\documentclass[aps,prd,onecolumn,superscriptaddress,floatfix,nofootinbib,showpacs,altaffilletter]{revtex4-2}

\usepackage{epsfig,graphicx,amssymb,amsmath,wasysym,graphicx}

\usepackage{amsmath}
\usepackage{amsfonts}
\usepackage{amssymb}
\usepackage{graphicx}
\usepackage{bm}
\usepackage{subfigure}
\usepackage{url}
\usepackage[hyperindex]{hyperref}
\usepackage{color}
\usepackage[ddmmyy,24hr]{datetime}
\usepackage{bigdelim}
\usepackage{booktabs}
\usepackage{dcolumn}
\usepackage{multirow}
\usepackage{subfigure}
\usepackage{cancel}
\usepackage{stackrel}
\usepackage{paralist}
\usepackage{xspace}
\usepackage{slashed}
\usepackage{cancel}
\usepackage{todonotes}
\usepackage{enumerate}
\usepackage{float}
\usepackage{fullpage}
\usepackage{ulem}
\usepackage{physics}

\newcommand{\nua}[1]{\ensuremath{\rlap{\kern-2.5pt\ensuremath{\overset{\scriptscriptstyle(-)}{\phantom{\nu}}}}{\ensuremath{{\nu}_{#1}}}}}
\newcommand{\vet}[1]{\ensuremath{\hskip-1pt\vec{\hskip1pt#1}}}
\usepackage{enumitem}

\newcommand{\cenns}{CE$\nu$NS\xspace}

\newcommand{\be}{\begin{equation}}
\newcommand{\ee}{\end{equation}}
\newcommand{\ba}{\begin{array}}
\newcommand{\ea}{\end{array}}

\usepackage{orcidlink}

\begin{document}

\title{When backgrounds become signals: neutrino interactions in xenon-based dark matter detectors}

\author{M. Atzori Corona \orcidlink{0000-0001-5092-3602}}
\email{mcorona@roma2.infn.it}
\affiliation{Istituto Nazionale di Fisica Nucleare (INFN), Sezione di Roma Tor Vergata, Via della Ricerca Scientifica, I-00133 Rome, Italy}
	
\author{M. Cadeddu \orcidlink{0000-0002-3974-1995}}
\email{matteo.cadeddu@ca.infn.it}
\affiliation{Istituto Nazionale di Fisica Nucleare (INFN), Sezione di Cagliari,
	Complesso Universitario di Monserrato - S.P. per Sestu Km 0.700,
	09042 Monserrato (Cagliari), Italy}

\author{N. Cargioli \orcidlink{0000-0002-6515-5850}}
\email{nicola.cargioli@ca.infn.it}
\affiliation{Istituto Nazionale di Fisica Nucleare (INFN), Sezione di Cagliari,
	Complesso Universitario di Monserrato - S.P. per Sestu Km 0.700,
	09042 Monserrato (Cagliari), Italy}

\author{F. Dordei \orcidlink{0000-0002-2571-5067}}
\email{francesca.dordei@cern.ch}
\affiliation{Istituto Nazionale di Fisica Nucleare (INFN), Sezione di Cagliari,
	Complesso Universitario di Monserrato - S.P. per Sestu Km 0.700,
	09042 Monserrato (Cagliari), Italy}
    
\author{M. Sestu \orcidlink{0009-0002-6044-7800
}}
\email{michela.sestu@ca.infn.it}
\affiliation{Istituto Nazionale di Fisica Nucleare (INFN), Sezione di Cagliari,
	Complesso Universitario di Monserrato - S.P. per Sestu Km 0.700,
	09042 Monserrato (Cagliari), Italy}
\affiliation{Dipartimento di Fisica, Universit\`{a} degli Studi di Cagliari,
	Complesso Universitario di Monserrato - S.P. per Sestu Km 0.700,
	09042 Monserrato (Cagliari), Italy}

\begin{abstract}
Direct detection dark matter experiments have proven to be compelling probes for studying low-energy neutrino interactions with both nuclei and atomic electrons, offering complementary information to accelerator and reactor-based neutrino experiments. Recently, the XENONnT and PandaX-4T collaborations reported the first evidence of coherent elastic neutrino-nucleus scattering from $^8\mathrm{B}$ solar neutrinos.
Thanks to their excellent background rejection capabilities and distinctive signal signatures, dual-phase time projection chambers are also sensitive to $pp$ solar neutrinos via their elastic scattering off atomic electrons in the target material. Although this signal is subdominant within the Standard Model, it becomes significantly enhanced in many beyond the Standard Model scenarios, offering a unique opportunity to probe new physics in the low-energy regime.
In this work, we analyze the latest electron recoil and nuclear recoil data from XENONnT, \mbox{PandaX-4T}, and LUX-ZEPLIN to probe Standard Model and Beyond the Standard Model physics.
While the precision of current neutrino measurements from such detectors remains lower than that achieved by dedicated neutrino experiments, their sensitivity to the tau neutrino component of solar neutrinos helps complete the overall picture, especially when investigating flavor-dependent new physics effects.
\end{abstract}

\maketitle

\section{Introduction}
Dual-phase time projection chambers (TPCs) filled with noble elements such as argon (Ar) or xenon (Xe) are widely used in the direct detection of particle dark matter searches~\cite{XENON:2024wpa, PandaX:2018wtu, LZ:2019sgr,DarkSide-20k:2024yfq,DarkSide-20k:2017zyg}. The universe, including our solar system, is believed to be permeated by a vast abundance of dark matter particles, which may interact with ordinary matter via weak-scale cross sections, for which reason they are often referred to as weakly-interacting massive particles (WIMPs). Due to their large target masses and excellent background discrimination, dual-phase TPCs are well suited to identifying rare and faint signals that arise from such interactions~\cite{Baudis:2023pzu}.
These detectors typically consist of a large volume of noble liquid, which serves as the primary active medium, and a thin gaseous phase situated above it. The signal is recorded by arrays of photosensors positioned at both the bottom and top of the chamber.
When a particle interacts within the liquid phase, it produces two distinct signals: a prompt scintillation signal (S1) and a delayed ionization signal (S2). The S1 signal originates from the prompt scintillation light emitted by the recoiling target atoms. The S2 signal is produced when the ionized electrons, freed during the interaction, are drifted upward by an applied electric field in the gas phase. There, they generate secondary scintillation light via electroluminescence. The combined measurement of S1 and S2 enables accurate event reconstruction and powerful background discrimination.
However, one of the most significant sources of background at low energies in dark matter searches arises from solar neutrinos  scattering elastically off nuclei (CE$\nu$NS) and electrons ($\nu$ES) in the target volume. Their signals can closely resemble those expected from dark matter particles, making them difficult to distinguish. Although traditionally regarded as a limitation in dark matter searches, the ability to detect solar neutrinos with existing dual-phase TPCs offers a compelling opportunity to study low-energy neutrino interactions. In principle, the electron recoil (ER) background can be reduced in the event selection procedure, however, in practice, it is difficult to fully eliminate it in xenon-based experiments due to the limited discrimination between nuclear recoils (NRs) and ERs. Fortunately, within the Standard Model (SM), the $\nu$ES contribution to the low-energy recoil spectrum is well understood and nearly flat with respect to recoil energy, allowing it to be subtracted in standard dark matter analyses.
However, in certain scenarios beyond the Standard Model (BSM), the $\nu$ES rate could be significantly enhanced, making its investigation particularly relevant. In contrast, the nuclear recoil signal induced by CE$\nu$NS is irreducible, as it closely mimics the signature expected from a potential WIMP interaction in the detector. Given the significant improvements in background rejection achieved by current detectors, dual-phase TPCs can now be exploited to study solar neutrinos in the absence of a dark matter signal, as we will discuss in this work.

\section{Theoretical framework}
\label{sec:cs}

\subsection{Coherent elastic neutrino-nucleus scattering}
Coherent elastic neutrino-nucleus scattering is a neutral current process in which a low-energy neutrino interacts with an atomic nucleus, which responds coherently. The interaction results in a small energy deposit due to the nuclear recoil, which makes it hard to detect. However, because of the coherent nuclear response, the cross section is enhanced and roughly scales with the square of the number of neutrons, making it relatively large compared to other neutrino interactions at low energy. The \cenns differential cross section as a function of nuclear recoil energy \(T_\mathrm{nr}\) for a neutrino \(\nu_\ell\) (\(\ell = e, \mu, \tau\)) scattering off a nucleus \(\mathcal{N}\) is given by~\cite{Cadeddu:2023tkp}
\begin{equation}
    \dfrac{d\sigma_{\nu_{\ell}\text{-}\mathcal{N}}}{d T_\mathrm{nr}} = 
    \dfrac{G_{\text{F}}^2 M}{\pi} 
    \left( 1 - \dfrac{M T_\mathrm{nr}}{2 E_\nu^2} \right)
    \left( Q^{V}_{\ell, \mathrm{SM}} \right)^2,
    \label{eq:cexsec}
\end{equation}
where \(G_{\text{F}}\) is the Fermi constant, \(E_\nu\) is the neutrino energy and \(M\) is the mass of the target nucleus. The strength of the interaction is quantified by the so-called nuclear weak charge, which is defined by
\begin{equation}
    Q^{V}_{\ell, \mathrm{SM}} = \left[ g_{V}^{p}(\nu_\ell) Z F_Z(|\vec{q}|^2) + g_{V}^{n} N F_N(|\vec{q}|^2) \right],
    \label{eq:weakcharge}
\end{equation}
where $g_{V}^{n}$ and $g_{V}^{p}(\nu_\ell)$ quantify the weak neutral-current interactions of neutrons and protons, respectively. In the SM, they correspond to 
\begin{align}
g_{V}^{p}(\nu_{e}) = 0.0379,\quad g_{V}^{p}(\nu_{\mu}) = 0.0297, \quad g_{V}^{p}(\nu_{\tau}) = 0.0253,\quad g_{V}^{n} = -0.5117, 
\end{align}
when taking into account radiative corrections~\cite{AtzoriCorona:2024rtv,AtzoriCorona:2023ktl,Erler:2013xha,PhysRevD.110.030001}. Interestingly, $g_{V}^{p}(\nu_\ell)$ depends on the weak mixing angle, a key parameter of the electroweak theory, and brings the only flavor dependence in the cross section due to the neutrino charge radius~\cite{AtzoriCorona:2025xwr,AtzoriCorona:2024rtv,Cadeddu:2018dux,Bernabeu:2002pd,Bernabeu:2002nw}. Moreover, the nuclear weak charge depends on the proton and neutron number, \(Z\) and \(N\), respectively, and on the nuclear form factors, \(F_Z(|\vec{q}|^2)\), \(F_N(|\vec{q}|^2)\), which describe the loss of coherence as a function of the momentum transfer \( |\vec{q}| \)~\cite{AtzoriCorona:2023ktl} and thus incorporate the dependence on fundamental nuclear parameters such as the proton and neutron nuclear radii.
Given that xenon is usually found as a compound of different isotopes, we use $(Z, N)_{\mathrm{Xe}} = (54, (70,72,74,75,76,77,78,80,82))$,
considering the natural abundances $\textit{f}(^{124}\mathrm{Xe}) = 0.000952, \, \textit{f}(^{126}\mathrm{Xe}) =  0.000890,\, \textit{f}(^{128}\mathrm{Xe}) = 0.019102,\, \textit{f}(^{129}\mathrm{Xe}) = 0.264006,\, \textit{f}(^{130}\mathrm{Xe}) =0.040710,\,\textit{f}(^{131}\mathrm{Xe}) =0.212324 \,\textit{f}(^{132}\mathrm{Xe}) =0.269086,\, \textit{f}(^{134}\mathrm{Xe}) =0.104 357, $ \,and $ \textit{f}(^{136}\mathrm{Xe}) = 0.088 573$~\cite{BerglundWieser+2011+397+410}.\\
In the case of solar neutrino data, the nuclear form factors are essentially equal to unity as the principal contribution comes from the interaction of boron-8 ($^8$B) solar neutrinos, which have energies up to $E_\nu\lesssim16\, \mathrm{MeV}$. This simplifies the analyses, making the results independent of specific form factor parameterisations. In any case, we consider the Helm parameterization~\cite{Helm:1956zz} with the values for Xe proton radii extracted from spectroscopy and electron scattering data~\cite{Fricke:1995zz, Angeli:2013epw, Fricke2004}, and the neutron radius estimated from theoretical shell models~\cite{Hoferichter:2020osn}. 

\subsection{Neutrino-electron elastic scattering}
The elastic scattering of neutrinos off atomic electrons, $\nu$ES, is a concurrent process to CE$\nu$NS. Although within the SM its contribution to the total event rate at low recoil energies is usually negligible, in BSM scenarios, the $\nu$ES contribution can increase significantly~\cite{Coloma:2022avw}.
Moreover, when dealing with solar neutrino data, the $\nu$ES contribution becomes even more relevant for the so-called low-energy analyses, where one searches for light dark matter particle (masses around the MeV scale). To look for such a tiny energy deposit one loses the discrimination power characteristic of dual-phase TPCs, introducing in the region-of-interest (ROI) events due to electron recoils, such as the one due to the elastic scattering of neutrinos off atomic electrons. 
The SM cross section for neutrino-electron elastic scattering off an atom \(\mathcal{A}\) is~\cite{Giunti:2014ixa}
\begin{align}
    \dfrac{d\sigma_{\nu_{\ell}-\mathcal{A}}}{d T_{\text{e}}}
    =
    Z_{\text{eff}}^{\mathcal{A}}
    \,
    \dfrac{G_{\text{F}}^2 m_{e}}{2\pi}
    \left[ \vphantom{\left( 1 - \dfrac{T_{e}}{E} \right)^2}
    \left( g_{V}^{\nu_{\ell}} + g_{A}^{\nu_{\ell}} \right)^2
    +
    \left( g_{V}^{\nu_{\ell}} - g_{A}^{\nu_{\ell}} \right)^2
    \left( 1 - \dfrac{T_{e}}{E_\nu} \right)^2
    -
    \left( (g_{V}^{\nu_{\ell}})^2 - (g_{A}^{\nu_{\ell}})^2 \right)
    \dfrac{m_{e} T_{e}}{E_\nu^2}
    \right],
    \label{eq:ES-cross-section}
\end{align}
where \(m_e\) is the electron mass, \(T_e\) is the electron recoil energy, $Z_{\text{eff}}^{\mathcal{A}}$~\cite{Coloma:2022avw,AtzoriCorona:2022jeb,Thompsonxray,Kouzakov:2014lka} accounts for the number of electrons ionized at a given recoil energy, and the flavor-dependent couplings are
\begin{align}
    g_{V}^{\nu_{e}} = 0.9524, \quad g_{A}^{\nu_{e}} = 0.4938,\quad g_{V}^{\nu_{\mu}} = -0.0394, \quad g_{V}^{\nu_{\tau}}=-0.0350,\quad g_{A}^{\nu_{\mu,\tau}} = -0.5062, \label{eq:gvm}
\end{align}
when including also radiative corrections~\cite{Erler:2013xha}. Similarly to the neutrino-proton coupling discussed for CE$\nu$NS, also $g_V^{\nu_\ell}$ depends on the weak mixing angle and on the neutrino charge radius~\cite{AtzoriCorona:2025xwr}.
In the sub-keV regime relevant for solar neutrinos, the main contribution of solar neutrino $\nu$ES stems from the so-called $pp$ solar neutrinos, which reach energies up to $E_\nu\lesssim0.4$ MeV. In such regime, atomic effects may become significant, requiring corrections to $Z_{\text{eff}}^{\mathcal{A}}$. This can be achieved using \textit{ab initio} methods like the multiconfiguration relativistic random phase approximation (MCRRPA)~\cite{PhysRevA.25.634,PhysRevA.26.734,Chen:2013lba}, which better accounts for many-body dynamics, or in some beyond the standard model scenarios, the so-called equivalent photon approximation (EPA) can be employed~\cite{Chen:2014ypv,Hsieh:2019hug}. The latter links the ionization cross section to the photoabsorption one, and it is usually applied when considering possible neutrino electric charges~\cite{AtzoriCorona:2022jeb}.\\

Given that both the CE$\nu$NS and $\nu$ES cross sections are flavor dependent, the calculation of the total SM cross section needs to account for the different contributions from all neutrino flavors present in the solar neutrino flux and the neutrino oscillation mechanism. Therefore, the total cross section becomes
\begin{equation}
\dfrac{d\sigma_{\nu-X}}{d T_{\text{X}}}
(E_\nu,T_{\text{X}})=
P_{ee}(E_\nu)\dfrac{d\sigma_{\nu_{e}\!-\!X}}{d T_{\text{X}}}
+\sum_{f=\mu,\tau}P_{ef}(E_\nu)\dfrac{d\sigma_{\nu_{f}-X}}{d T_{\text{X}}}
,
\label{ES-SM}
\end{equation}
where  $T_{\text{X}}$ is either the NR ($T_{\rm nr}$) or ER ($T_e$) recoil energy and $P_{ee}(E_\nu)=\sin^4\theta_{13}+\cos^4\theta_{13}P^{2\nu}$~\cite{Borexino:2017fbd} is the survival probability for solar neutrinos reaching the detector, which is a function of the neutrino energy due to the MSW mechanism. The term $P^{2\nu}\simeq  0.55$~\cite{Chen_2017,PhysRevD.110.030001} refers to the $\nu_e$ survival probability in the two-neutrino oscillation scheme. Here, $P_{e\mu}=(1-P_{ee})\cos^2\theta_{23}$ and $P_{e\tau}=(1-P_{ee})\sin^2\theta_{23}$ are the transition probabilities. The values of the corresponding mixing angles $\theta_{13}$ and $\theta_{23}$ were taken from Ref.~\cite{PhysRevD.110.030001}.


\subsection{Data Analysis Strategy}
In this work, we have analyzed five different datasets coming from three different collaborations. Namely, both the NR and ER data from XENONnT (XnT)\footnote{We consider both the SR0 and SR1 NR data from the XENONnT experiment.}~\cite{XENON:2024ijk, XENON:2022ltv} and PandaX-4T (P4T)\footnote{We consider both the Run-0 and Run-1 ER data from the PandaX-4T experiment.}~\cite{PandaX:2024muv, PandaX:2024cic} and the ER data\footnote{During the completion of this work, the LZ collaboration also reported their first evidence for the \cenns signal with an improved statistical significance~\cite{LZ:2025igz}.} from Lux-Zeplin (LZ)~\cite{LZ:2024zvo}. For the analysis of the datasets, we employed the information provided by the collaborations in their publications and the accompanying data releases and supplemental materials unless noted otherwise.
To analyse the data, we have to first calculate the expected number of \cenns and $\nu$ES events. To do so, we have considered the total differential neutrino flux, $d N_{\nu,j}/d E_\nu$, given by the sum of all the different solar neutrino components $j$ as from Refs.~\cite{PhysRevD.110.030001, Vitagliano:2019yzm}. In particular, the most relevant contributions to $\nu$ES analyses are given by the continuous $pp$ flux and the monochromatic $^{7}$Be 861~keV line. Instead, for CE$\nu$NS searches, the main events come from the $^8$B neutrino flux together with a subdominant contribution from the \textit{hep} neutrinos at higher energies.
The expected theoretical CE$\nu$NS or $\nu$ES number of events in each energy bin $i$ can be retrieved as
\begin{align}\label{N_es}
N_i^{\rm \nu }
=
N(\mathrm{Xe})
\int_{T_{\mathrm{X}}^{i}}^{T_{\mathrm{X}}^{i+1}}
\hspace{-0.1cm}
d T_{\mathrm{X}}\,
A(T_{\mathrm{X}})\int_{E^{\rm min}_{\nu}(T_{\text{X}})}^{E^{\rm max}_{\nu}}
\hspace{-0.3cm}
d E_\nu
\sum_{j}
\frac{d N_{\nu,j}}{d E_\nu}(E_\nu)
\frac{d \sigma_{\nu\!-\!X}}{d T_{\mathrm{X}}}(E_\nu, T_{\mathrm{X}})
,
\end{align}
where $N(\mathrm{Xe})$ is the number of xenon targets contained in the detector,
$A(T_{\text{X}})$ is the energy-dependent detector efficiency and the minimum neutrino energy is set by kinematic as 
$E^{\rm min}_{\nu}(T_{\text{X}}) = (T_\text{X}+\sqrt{T_\text{X}^{2} + 2m_\text{X} T_\text{X}})/2$, where $m_\text{X} = M$ ($m_e$) for CE$\nu$NS ($\nu$ES). The maximum neutrino energy instead depends on the neutrino flux and is set to $E^{\rm max}_{\nu,\text{NR}} \sim 20$~MeV and $E^{\rm max}_{\nu,\text{ER}} \sim 2$~MeV.
The number of target Xe atoms in the detector is given by
$N(\mathrm{Xe}) = N_{\mathrm{A}} M_{\mathrm{det}} / M_{\mathrm{\mathrm{Xe}}}$, 
where $N_{\mathrm{A}}$ is the Avogadro number, 
$M_{\mathrm{det}}$ is the detector fiducial mass and
$M_{\mathrm{\mathrm{Xe}}}$ is the average xenon molar mass.
Given that the CE$\nu$NS data are provided not as a function of NR energy but in terms of observed charge of the S2 signal, we have converted such a scale in order to retrieve the corresponding nuclear recoil bin following closely the information provided by the XnT and P4T collaborations, and taking advantage of the provided values of the charge yields.\\
To compare the theoretical number of predicted events to the experimental data, we analyse ER data using a Poissonian least-squares function~\cite{Baker:1983tu,PhysRevD.110.030001} in order to properly treat some energy bins where the number of events is small, namely
\begin{align}
\label{eq:chi2xe}
    \chi^2_{\rm ER}
     \!=\!
    2\sum_{i}
    \left[
    (1+\alpha) N_i^{\rm bkg}
    + (1+\beta) N_i^{\rm \nu ES}
    - N_i^{\rm exp}+ N_i^{\rm exp} \ln\left(\frac{N_i^{\rm exp}}{(1+\alpha) N_i^{\rm bkg}
    + (1+\beta) N_i^{\rm \nu ES}}\right)
    \right]\!+\!\Big(\dfrac{\alpha}{\sigma_\alpha}\Big)^2
    \!+\!\Big(\dfrac{\beta}{\sigma_\beta}\Big)^2,
\end{align}
where $N_i^{\rm bkg}$ is the number of residual background events found in the $i$th bin, $N_i^{\rm \nu ES}$ is the prediction in the $i$th bin for the $\nu$ES signal, and $N_i^{\rm exp}$ is the experimental number of events in the $i$th bin.
We consider a single nuisance parameter, $\alpha$, to account for the uncertainty on the background component. This procedure ignores the fact that the different background contributions have different relative uncertainties; however, given that the total background is dominated by the $\beta$ decays this approximation is valid. In particular we consider $\sigma_\alpha=({0.038,0.018,0.027})$ for LZ, XnT and P4T, respectively.
Instead, the nuisance parameter $\beta$ takes into account the uncertainty on the neutrino flux (with $\sigma_\beta=7\%$)\footnote{The flux uncertainty is about 7\% for $^7$Be and 0.6\% for $pp$~\cite{Serenelli:2016dgz}, we conservatively use the first one for both fluxes.}. A similar approach is also used in Refs.~\cite{Demirci:2025poc,Maity:2024aji,AtzoriCorona:2022jeb}.\\
Instead, for the NR data we consider a Gaussian least-squares function in the form of 
\begin{equation}
    \chi^2_{\rm NR}=\sum_{i}\left(\dfrac{N_{i}^{\rm exp}-(1+\eta)N_{i}^{\rm CE\nu NS}-\sum_j(1+\rho_j)N_{i}^{\rm bkg^j}}{\sigma_{\rm exp}}\right)^2+\left(\dfrac{\eta}{\sigma_\eta}\right)^2+\sum_j \left(\dfrac{\rho_j}{\sigma_{\rho_j}}\right)^2\, ,\label{chi2-NR}
\end{equation}
where $N_i^{\rm bkg^j}$ is the number of $j$-type background events found in the $i$th bin, $N_i^{\rm CE\nu NS}$ is the prediction in the $i$th bin for the CE$\nu$NS signal, and $N_i^{\rm exp}$ is the experimental number of events in the $i$th bin with corresponding uncertainty $\sigma_\mathrm{exp}$. The nuisance parameters $\rho_j$ take into account the uncertainty on the $j$-type background, with the sum over $j$ running over all the background components considered independently. For XnT we consider the different contributions due to accidental coincidences, neutron and electron-recoil backgrounds with relative uncertainties \mbox{$\sigma_\rho=({0.09,0.54,1})$} and \mbox{$\sigma_\rho=({0.058,0.58,1})$} for the SR0 and SR1 datasets, respectively. For the P4T US2 data, instead we consider the contributions from the cathode and the micro-discharging backgrounds, with corresponding uncertainties $\sigma_\rho=({0.31,0.23})$, respectively.
The nuisance $\eta$ takes into account the uncertainty on the $^{8}$B neutrino flux, with $\sigma_\eta=7\%$~\cite{Borexino:2017uhp}.
A similar analysis framework for dealing with the XnT data has been considered in Refs.~\cite{DeRomeri:2024iaw,Maity:2024aji}, where SR0 and SR1 have not been considered separately, but as a total contribution.
On the contrary, the P4T collaboration reports the results from a joint analysis of US2 and paired data, but not enough information is available to reproduce the latter. Therefore, here we will report the results from a spectral analysis of the US2 data alone, which, when comparable, yields weaker bounds and constraints with respect to the official ones~\cite{PandaX:2024muv}. Moreover, to match their precision and show the full potentiality of such a measurement, we consider also an analysis based on the integrated number of CE$\nu$NS events from both the US2 and paired data in those physics scenarios where no significant profit from the spectral information is expected, i.e. for the flux normalization, the weak mixing angle and the neutrino charge radius, following a similar approach to that employed in Ref.~\cite{AristizabalSierra:2024nwf} for the study of neutrino non standard interactions.
Therefore, we consider a simple Gaussian least-squares function in the form of 
\begin{equation}
    \chi^2_{\rm US2+paired}=\left(\dfrac{N_{\rm US2}^{\rm exp}-(1+\eta)N_{\rm US2}^{\rm CE\nu NS}}{\sigma^{\rm US2}_{\rm exp}}\right)^2+\left(\dfrac{N_{\rm paired}^{\rm exp}-(1+\eta)N_{\rm paired}^{\rm CE\nu NS}}{\sigma^{\rm paired}_{\rm exp}}\right)^2+\left(\dfrac{\eta}{\sigma_\eta}\right)^2 ,
\end{equation}
where $N_{\rm US2}^{\rm exp}=75\pm28$ and $N_{\rm paired}^{\rm exp}=3.5\pm1.3$ are the total number of CE$\nu$NS events from the US2 and paired data respectively~\cite{PandaX:2024muv}, with the corresponding uncertainties. Instead, $\eta$ is the nuisance parameter on the theoretical number of events, with uncertainty $\sigma_\eta=7\%$, as done in the full spectral analysis in Eq.~(\ref{chi2-NR}). Such nuisance parameter is removed when fitting for the neutrino flux normalization.

\section{Results}

\subsection{Standard Model tests}
\begin{figure}[h]
    \centering
    \subfigure[]{\label{fig:etaCEvNS}
    \includegraphics[width=0.48\linewidth]{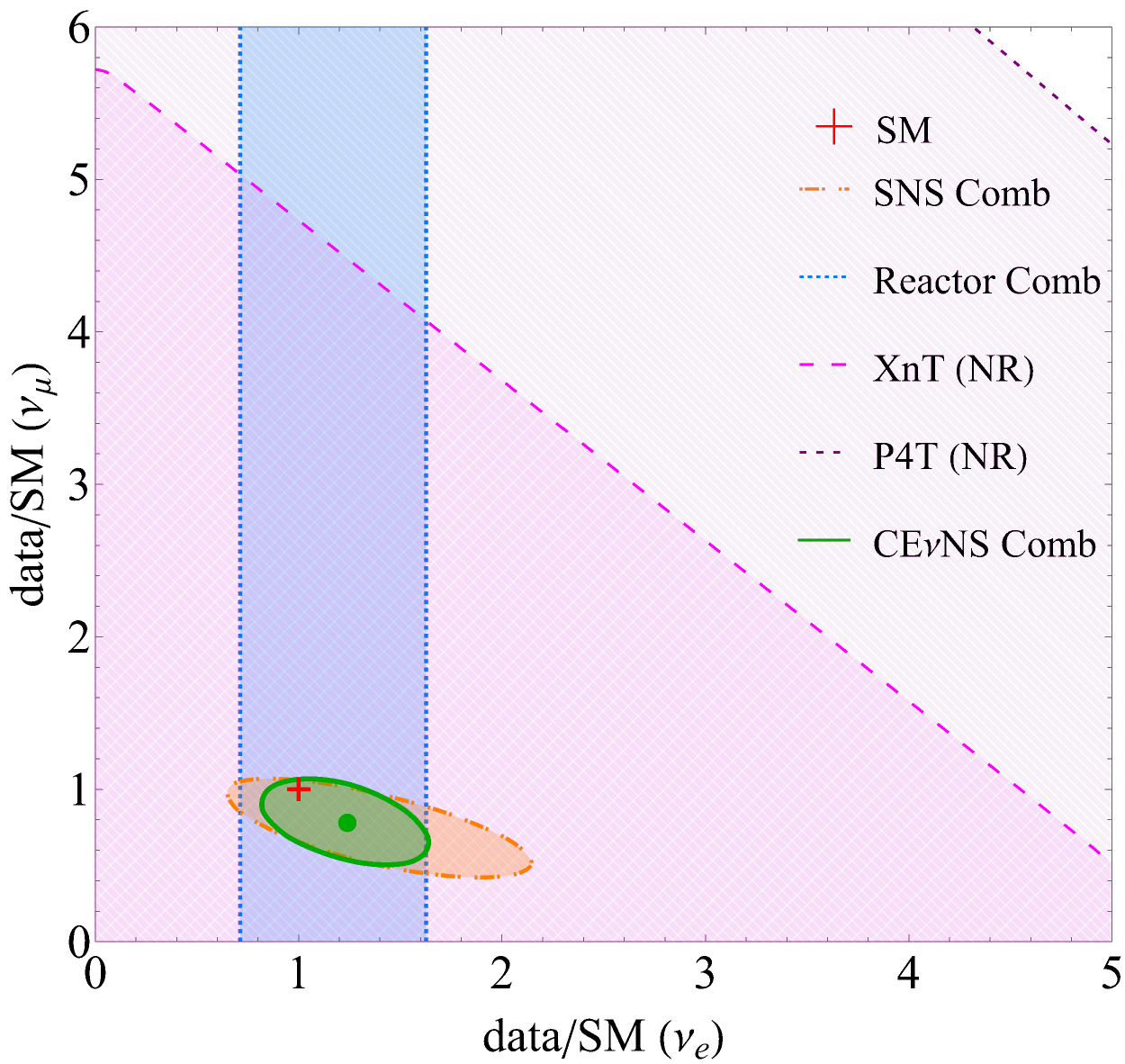}}
    \subfigure[]{\label{fig:etatau}
    \includegraphics[width=0.49\linewidth]{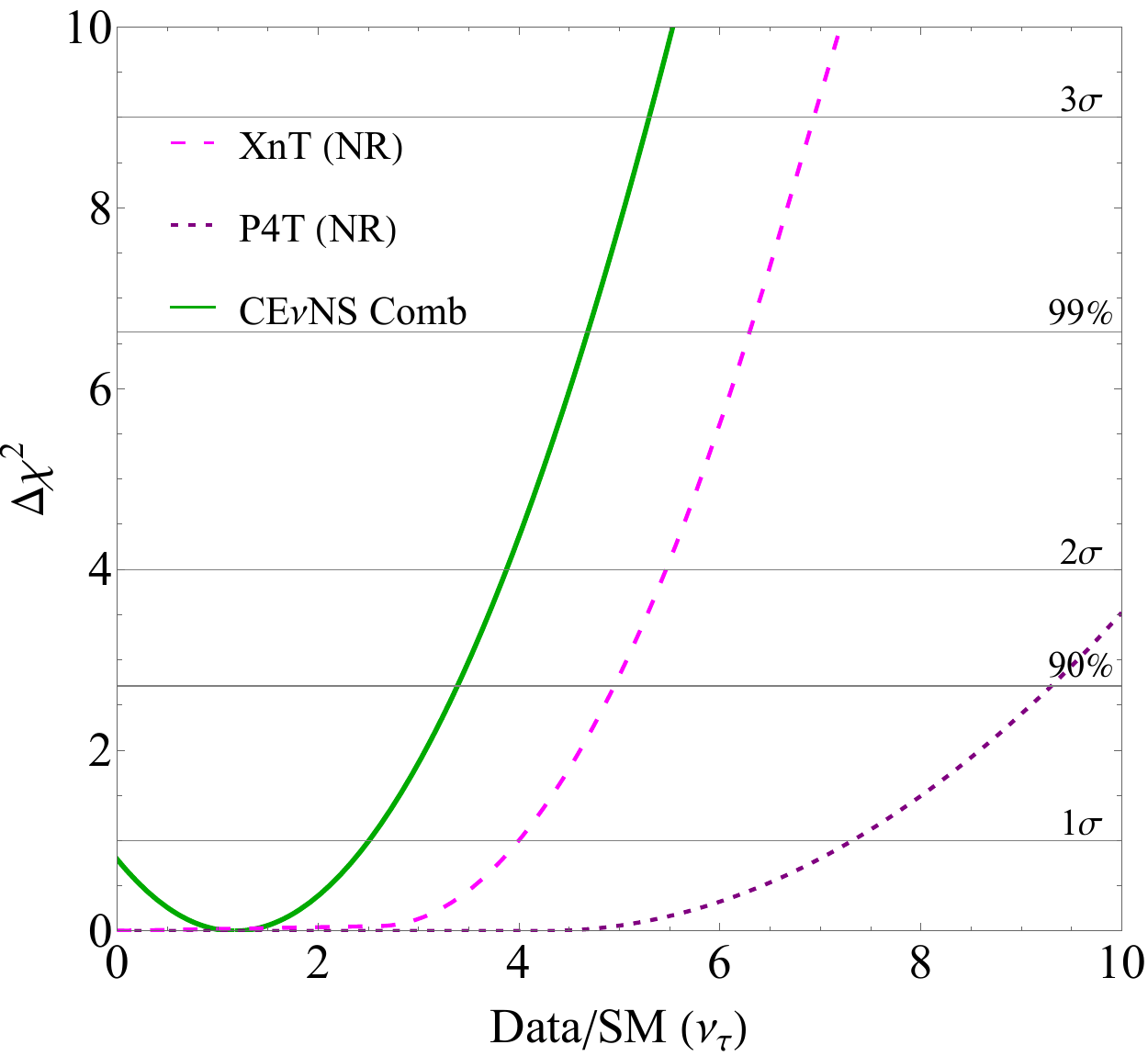}}
    \caption{Left: Agreement between CE$\nu$NS data and SM prediction in the muon and electron neutrino flavor basis at $1\sigma$ CL for the solar neutrino data compared to the most stringent spallation neutron source and reactor neutrino results~\cite{AtzoriCorona:2025xgj,atzoricorona:2025ygn}. Also shown is the total CE$\nu$NS combined fit. 
    Right: Marginal $\Delta\chi^2$ result on the tau neutrino flavor, comparing the solar neutrino CE$\nu$NS result to that from the combined fit.}\label{fig:eta}
\end{figure}

As a first test, we checked the consistency between the available data and theoretical predictions, finding a good overall agreement, at least within the current level of precision. This is shown in Fig.~\ref{fig:eta}, in terms of the ratio between CE$\nu$NS data and predictions separated for the different flavor components. We compare the results of this work obtained considering the least-square function in Eq.~(\ref{chi2-NR}), with those from the other available CE$\nu$NS measurements\footnote{A compilation of results from CE$\nu$NS and $\nu$ES probes can be found in the LE$\nu$S web page at \url{https://levs-fit.ca.infn.it}.}. The precision of current solar neutrino CE$\nu$NS data is still poor compared to the other available measurements~\cite{AtzoriCorona:2025xgj,atzoricorona:2025ygn}, however, it allows probing the agreement in the $\tau$ neutrino flavor basis, which would otherwise be inaccessible. Therefore, in Fig.~\ref{fig:etatau}, we show the result on the $\tau$ flavor from our analysis of solar CE$\nu$NS together with the result of the joint analysis of all the available data. By exploiting the precision on the $\mu$ and $e$ flavors set by the other datasets, we can significantly improve the constraints on the $\tau$ flavor, finding a good agreement between data and theory, numerically corresponding to $\text{Data/SM}~(\nu_\tau) < 2.5$ at $1\sigma$ CL. \\
Moreover, the current precision allows us to put meaningful constraints on different SM and BSM scenarios, which will be analyzed in the following. 
\begin{figure}[h]
    \centering
    \subfigure[]{\label{fig:flux}
    \includegraphics[width=0.48\linewidth]{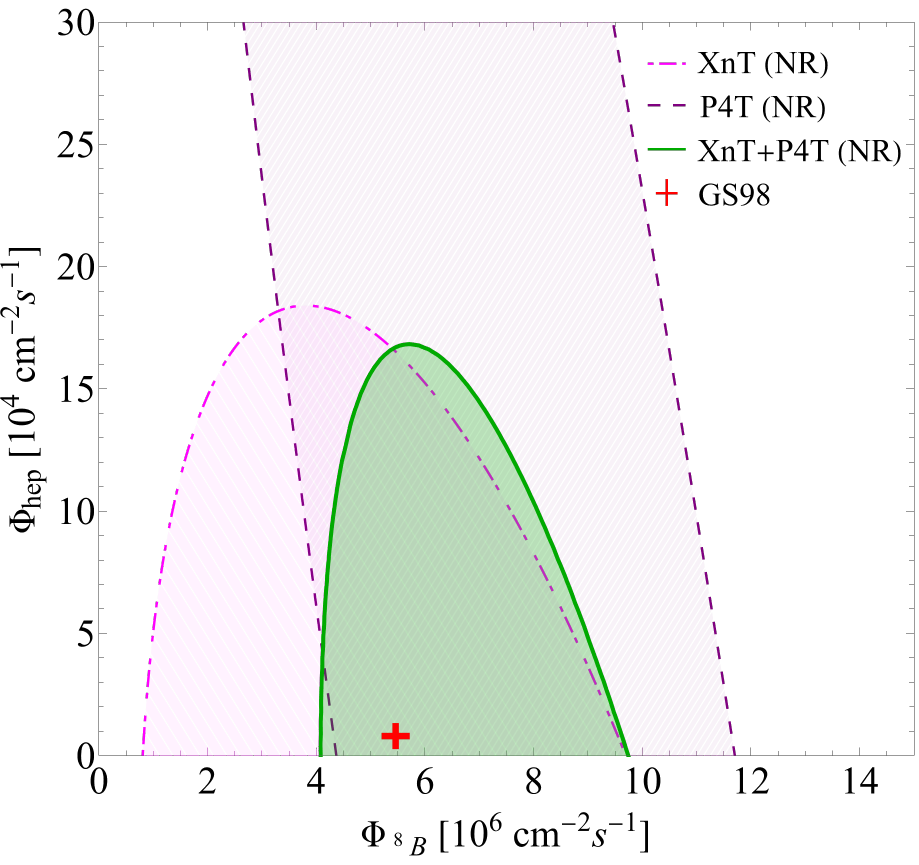}}
    \subfigure[]{\label{fig:fluxcomp}
    \includegraphics[width=0.5\linewidth]{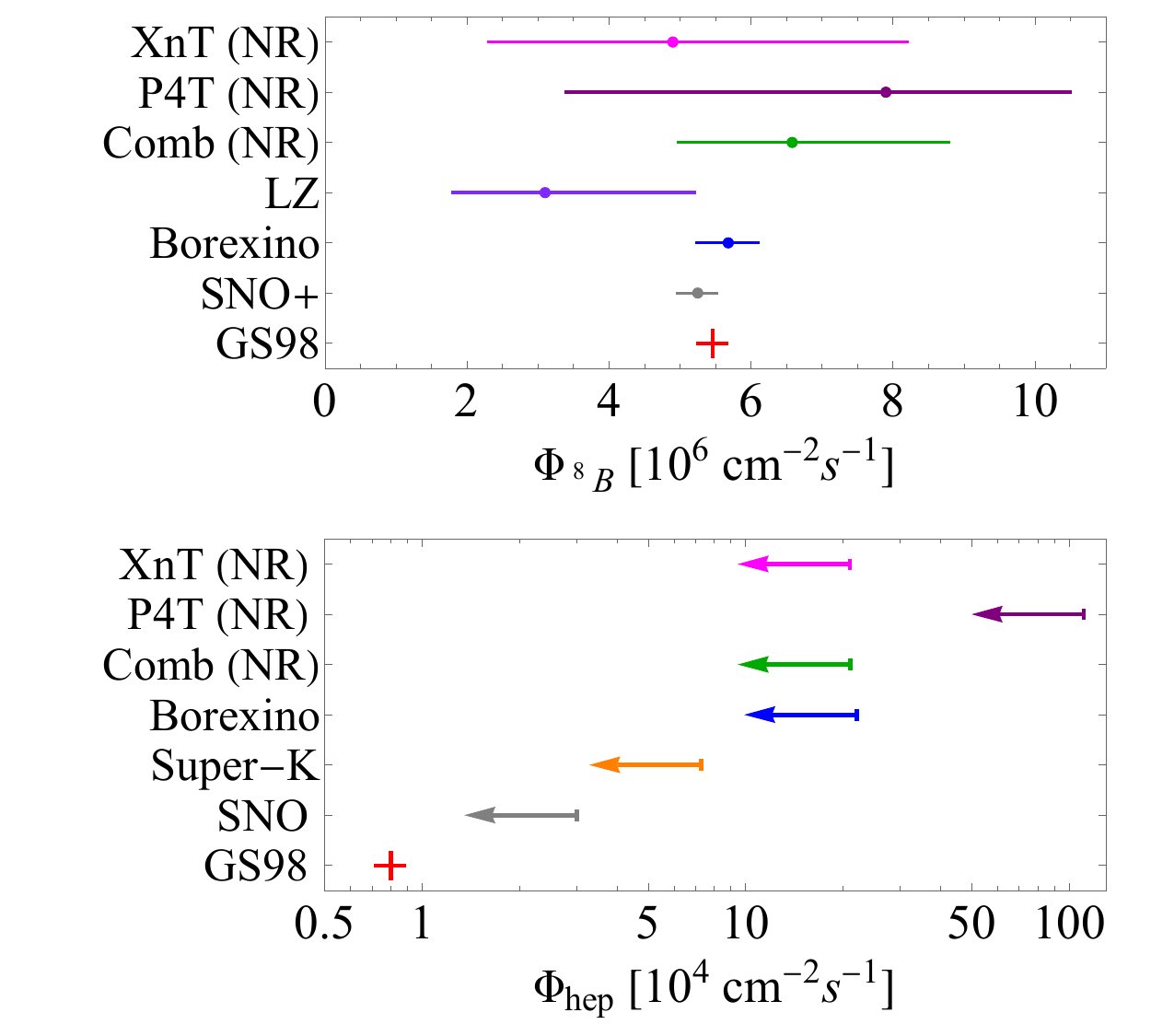}}
    \caption{Left: Solar neutrino flux normalization measurements at $1\sigma$ CL from the XnT and P4T data, together with their combination compared to the prediction from the GS98 solar model~\cite{Vitagliano:2019yzm}. 
    Right: The $1\sigma$ CL results for the $^8\mathrm{B}$ (top) and 90\% CL limits for the \textit{hep} (bottom) solar neutrino fluxes. Also shown are the constraints on the $^8\mathrm{B}$ flux from the LZ~\cite{LZ:2025igz}, Borexino~\cite{BOREXINO:2018ohr} and SNO+~\cite{SNO:2018fch} experiments and the constraints on the \textit{hep} flux from the Borexino~\cite{BOREXINO:2018ohr}, Super-K~\cite{Super-Kamiokande:2005wtt} and SNO~\cite{SNO:2020gqd} experiments.}
    \label{fig:fluxes}
\end{figure}

\begin{figure}
    \centering
    \subfigure[]{\label{fig:running}
    \includegraphics[width=0.5\linewidth]{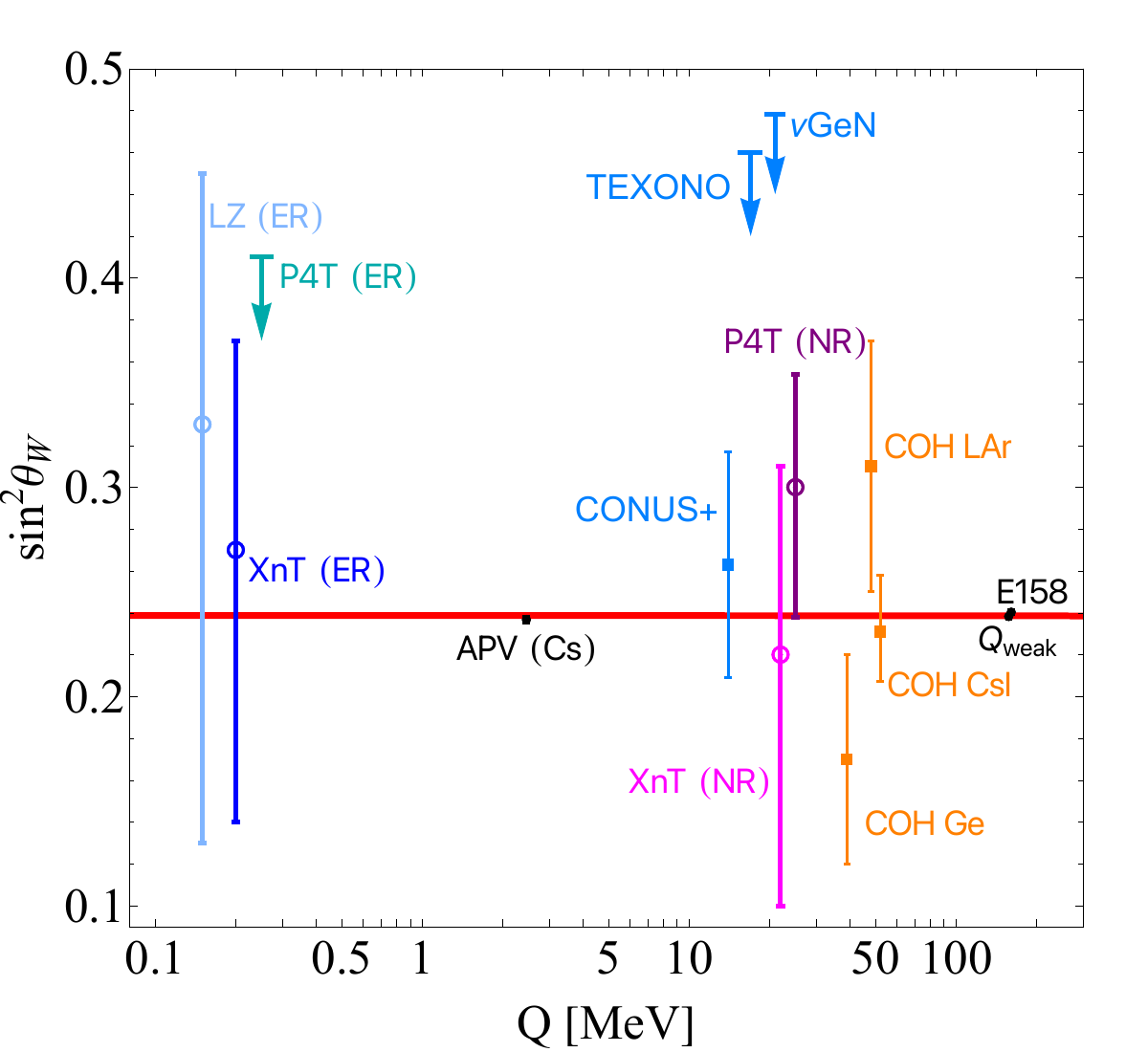}}
    \subfigure[]{\label{fig:NCRtau}
    \includegraphics[width=0.47\linewidth]{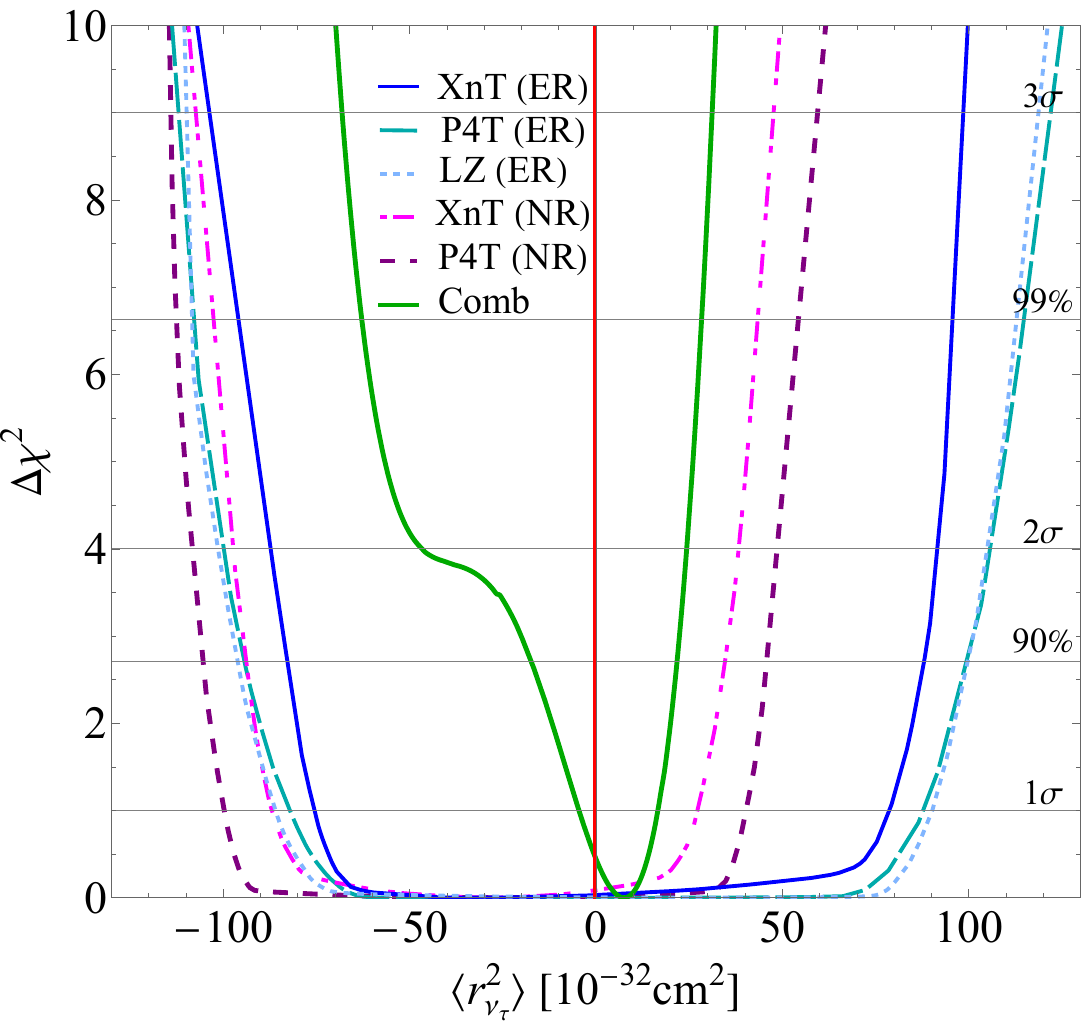}}    
    \caption{Left: Measurements of the weak mixing angle at low energy from solar neutrino data, both through ER and NR, compared with other available constraints~\cite{atzoricorona:2025ygn,AtzoriCorona:2025xgj,WoodAPV,Dzuba:2012kx,Anthony:2005pm,Androic:2018kni}. The red line depicts the SM prediction for the weak mixing angle~\cite{PhysRevD.110.030001}. The arrows refer to upper limits from the dataset which show less constraining power. Right: Marginal $\Delta\chi^2$ for the $\tau$ neutrino charge radius from each solar neutrino dataset analysed in this work, compared to the updated result of a global fit of neutrino-electron and \cenns data (green curve)~\cite{AtzoriCorona:2025xwr}.}
\end{figure}

A similar check of the overall agreement of CE$\nu$NS data can be obtained in terms of a measurement of the normalization of the involved neutrino fluxes. For \cenns data, the main contribution stems from $^8\mathrm{B}$ solar neutrinos, however, also a small contribution from solar \textit{hep} neutrinos may be observable. Therefore, we can constrain such normalizations by fitting the data fixing the cross section to its SM prediction. Given that  \textit{hep} neutrinos contribute only marginally, we expect more sensitivity on the $^8\mathrm{B}$ flux normalization. This is confirmed by the results shown in Fig.~\ref{fig:fluxes}, where we compare the results from the P4T and XnT data with their combination. 
In particular, P4T is practically insensitive to the \textit{hep} contribution, while XnT shows a better sensitivity thanks to the spectral information considered in the analysis, yielding to
\begin{align}
    &\Phi_{^{8}\mathrm{B}}(\text{XnT})=5.0^{+3.2}_{-2.7} \times 10^{6}\, \mathrm{cm}^{-2} \mathrm{s}^{-1} (1\sigma\, \mathrm{CL})\, ,\quad\quad\,\,\,\,\, \Phi_{\mathrm{hep}}(\text{XnT})<21 \times 10^{4}\, \mathrm{cm}^{-2} \mathrm{s}^{-1} (90\%\, \mathrm{CL})\,,\\
    &\Phi_{^{8}\mathrm{B}}(\text{P4T})=7.9^{+2.6}_{-4.5} \times 10^{6}\, \mathrm{cm}^{-2} \mathrm{s}^{-1}(1\sigma\, \mathrm{CL})\, ,\quad\quad\quad \Phi_{\mathrm{hep}}(\text{P4T})<11 \times 10^{5}\, \mathrm{cm}^{-2} \mathrm{s}^{-1}(90\%\, \mathrm{CL})\,,\\
    &\Phi_{^{8}\mathrm{B}}(\text{XnT+P4T})=6.7^{+2.1}_{-1.7} \times 10^{6}\, \mathrm{cm}^{-2} \mathrm{s}^{-1} (1\sigma\, \mathrm{CL})\,,\, \Phi_{\mathrm{hep}}(\text{XnT+P4T})<19 \times 10^{4}\, \mathrm{cm}^{-2} \mathrm{s}^{-1} (90\%\, \mathrm{CL})\,,
\end{align}
which are in agreement within uncertainty with the expectations from the GS98 solar model~\cite{Vitagliano:2019yzm}, which predicts \mbox{$\Phi_{^{8}\text{B}}=5.46\times 10^{6}\, \mathrm{cm}^{-2} \mathrm{s}^{-1}$} and \mbox{$\Phi_{\text{hep}}=0.8\times 10^{4}\, \mathrm{cm}^{-2} \mathrm{s}^{-1}$}, as depicted by the red cross in Fig.~\ref{fig:fluxes}. 
Given that for PandaX-4T we consider an analysis based on the total number of measured events, the actual constraint should be set on the sum of the two fluxes, leading to $\Phi_{^{8}\mathrm{B}+\text{hep}}(\text{P4T})=8.0^{+2.4}_{-2.4} \times 10^{6}\, \mathrm{cm}^{-2} \mathrm{s}^{-1}$ at $1\sigma$ CL.
Our result is also in perfect agreement with those from the XENONnT~\cite{XENON:2024ijk} and PandaX-4T~\cite{PandaX:2024muv} collaborations. Moreover, the result from our spectral analysis of P4T US2 data, leads to \mbox{$\Phi_{^{8}\mathrm{B}}(\text{P4T})=8.5^{+6.2}_{-6.1} \times 10^{6}\, \mathrm{cm}^{-2} \mathrm{s}^{-1}$}, which is in good agreement with the non-spectral result, although with a much worse precision. A comparable result has been reported in Ref.~\cite{DeRomeri:2024iaw} using a similar data analysis strategy. Although current data only allow us to set limits on the \textit{hep} flux normalization, this represents one of the few available constraints for such a flux.
To put this in context, in Fig.~\ref{fig:fluxcomp}, we report other available constraints. Namely, the measurements of the $^8\mathrm{B}$ flux from the LZ~\cite{LZ:2025igz}, Borexino~\cite{BOREXINO:2018ohr} and SNO+~\cite{SNO:2018fch} experiments and the constraints on the \textit{hep} flux from the Borexino~\cite{BOREXINO:2018ohr}, Super-K~\cite{Super-Kamiokande:2005wtt} and SNO~\cite{SNO:2020gqd} experiments. It is interesting to note that our combined analysis from the first solar neutrino data from DM detectors already yields a constraint on the \textit{hep} flux normalization at the level of that from Borexino~\cite{BOREXINO:2018ohr}, and only one order of magnitude weaker than the current world-best constraint by SNO~\cite{SNO:2020gqd}, namely \mbox{$\Phi_{\mathrm{hep}}(\text{SNO})<3 \times 10^{4}\, \mathrm{cm}^{-2} \mathrm{s}^{-1}$} at 90\% CL.
Thus, dark matter detectors show intriguing capability for measuring solar neutrino fluxes and in particular constraining the \textit{hep} neutrino flux normalization, which would result in a deeper understanding of solar dynamics.
This was also shown in Ref.~\cite{Xia:2024ytb}, where a sensitivity study for the measurement of solar neutrino fluxes with dark matter detector has been performed, showing that, with improved backgrounds and systematics, a precision of the order of $\sim 10\%$ on the $^{8}$B flux normalization may be reached, which could even improve up to $\sim3\%$ with increased exposures.
\\ 
On the contrary, relying on the GS98 solar model, the data can be used to constrain the weak mixing angle, $\theta_{\text{W}}$, a key parameter of the electroweak theory. Such parameter has been measured across various energy scales~\cite{PhysRevD.110.030001}, to test the prediction of the SM, which can be significantly altered in certain BSM scenarios~\cite{Safronova_2018,Cadeddu:2021dqx,Cadeddu:2024baq}.
The low-energy regime is still to be precisely tested, as only few probes are available. Among them, \cenns and $\nu$ES provide many determinations. In Fig.~\ref{fig:running} we show the results obtained in this work, along with other available measurements at low energies, such as those from other CE$\nu$NS experiments~\cite{atzoricorona:2025ygn,AtzoriCorona:2025xgj}, that from atomic parity violation on cesium~\cite{WoodAPV,Dzuba:2012kx} and the determinations of electron and proton nuclear weak charges~\cite{Androic:2018kni,Anthony:2005pm}.
We find no deviations from the SM prediction, although the uncertainties are still large. However, it is worth remarking that these are the only determinations of the weak mixing angle from solar neutrino data with dark matter experiments, and thus are expected to improve significantly in the future~\cite{Maity:2024aji,AtzoriCorona:2025xwr,Giunti:2023yha}.
Moreover, as pointed out in Ref.~\cite{Maity:2024aji}, solar neutrino ER data provide the weak mixing angle measurement at the lowest available energy, in a completely new regime.
Here, we report the numerical results obtained in this work
\begin{align}
\sin^2\theta_W(\mathrm{XnT-ER})=0.27^{+0.10}_{-0.13}\, ,&\quad
\sin^2\theta_W(\mathrm{XnT-NR})=0.22^{+0.09}_{-0.12}\, , \\ 
\sin^2\theta_W(\mathrm{P4T-ER})<0.41\, , &\quad
\sin^2\theta_W(\mathrm{P4T-NR})=0.30^{+0.05}_{-0.06}\, ,\\ 
\sin^2\theta_W(\mathrm{LZ-ER})=0.33^{+0.12}_{-0.20}\, ,&
\end{align}
where ER refers to $\nu$ES data, while NR to the CE$\nu$NS ones. The spectral analysis of P4T data yields a much weaker constraint, namely \mbox{$\sin^2\theta_W(\mathrm{P4T-NR})=0.32^{+0.14}_{-0.20}$}, which is in excellent agreement with Ref.~\cite{DeRomeri:2024iaw}, where a similar data analysis strategy is followed. Our result from XnT NR is also in good agreement with that of Ref.~\cite{DeRomeri:2024iaw}. Moreover, in Ref.~\cite{Maity:2024aji}, the authors performed a similar analysis of P4T and XnT NR data as well as XnT ER ones. They find a result in perfect agreement with ours concerning the ER data, while constraints only marginally compatible for the NR ones.

A deviation of the weak mixing angle with respect to its SM prediction could be reinterpreted in terms of the neutrino charge radius (CR), the only nonzero electromagnetic properties of neutrinos, usually discussed as radiative corrections to $g_{V}^{p}(\nu_{\ell})$ and $g_V^{\nu_\ell}$~\cite{Giunti:2024gec}. The SM prediction for the neutrino charge radii is given by~\cite{Bernabeu:2000hf,Bernabeu:2002nw}
\begin{equation}
\langle{r}_{\nu_{\ell}}^2\rangle_{\text{SM}}
=
-
\frac{G_{\text{F}}}{2\sqrt{2}\pi^2}
\left[
3 - 2 \ln\left(\frac{m_{\ell}^2}{m_W^2}\right)
\right],
\label{eq:cr-sm}
\end{equation}
where $m_{W}$ is the $W$ boson mass and $m_{\ell}$ is the mass of the charged lepton $\ell = e, \mu, \tau$, and numerically corresponds to
\begin{align}
    \langle r^2_{\nu_e}\rangle \simeq-0.83\times 10^{-32}\, \mathrm{cm}^2\, ,\quad
    \langle r^2_{\nu_\mu}\rangle \simeq-0.48\times 10^{-32}\, \mathrm{cm}^2\, ,\quad
    \langle r^2_{\nu_\tau}\rangle \simeq-0.30\times 10^{-32}\, \mathrm{cm}^2\, . \label{SMvaluesNCR}
\end{align}
The neutrino charge radius depends on the flavor of the neutrino and generates a photon mediated interaction with the neutrino at loop level. In the SM, such an interaction conserves the neutrino flavor, so that only diagonal terms are present. However, in some beyond the SM scenarios, also off-diagonal terms may be considered, although expected to be much smaller than the diagonal contribution, and are usually referred to as transition CR, $\langle r_{\nu_{\ell\ell'}}^2 \rangle$~\cite{Kouzakov:2017hbc,Cadeddu:2018dux}. In this work, we will only consider the scenario including diagonal terms. \\
To make explicit the dependence of the \cenns cross section on the neutrino charge radii, we can substitute the neutrino-proton coupling in Eq.~(\ref{eq:weakcharge}) with $g_V^p= \tilde{g}_{V}^{p} - \tilde{Q}_{\ell}$,
where $\tilde{g}_{V}^{p}=0.0182$ is the flavor independent neutrino-proton coupling and $\tilde{Q}_{\ell}$ represents the CR radiative contribution
\begin{equation}
\tilde{Q}_{\ell} = \dfrac{\sqrt{2} \pi \alpha}{3 G_{\text{F}}} \langle r_{\nu_{\ell}}^2 \rangle,
\label{eq:cr-term}
\end{equation}
with $\alpha$ being the fine-structure constant. Similarly, in the case of the $\nu$ES process, it is sufficient to substitute the neutrino-electron vector coupling in Eq.~(\ref{eq:ES-cross-section}) with $g_V^{\nu_\ell}= \tilde{g}_{V}^{\nu} + \tilde{Q}_{\ell}$,
where $\tilde{g}_{V}^{\nu}=-0.0279+\delta_{\ell\,e}$, with $\delta_{\ell\,e}$ accounting for the charge current contribution to the $\nu_e-e$ scattering. Further corrections related to the experimental momentum transfer have been included following Refs.~\cite{AtzoriCorona:2024rtv,AtzoriCorona:2025xwr}.\\
Solar neutrino data allow one to set constraints on all three neutrino flavors, although with a precision not yet competitive with that of other available CE$\nu$NS and $\nu$ES measurements~\cite{Akimov:2021dab,COHERENT:2020iec,COHERENT:2020ybo,Ackermann:2025obx,TEXONO:2024vfk,nuGeN:2025mla,TEXONO:2009knm,LSND:2001akn,Allen:1992qe,Ahrens:1990fp,CHARM-II:1993phx,CHARM-II:1994dzw,CHARM-II:1994aeb,CCFR:1997zzq}, which are instead sensitive only to $\langle r^2_{\nu_e}\rangle$ and $\langle r^2_{\nu_\mu}\rangle$.  
Therefore, we focus here on the results for $\langle r^2_{\nu_\tau}\rangle$, which are shown in Fig.~\ref{fig:NCRtau} for each individual dataset after marginalizing over the remaining two flavor charge radii.
Numerically, the 90\% CL bounds on $\langle r^2_{\nu_\tau}\rangle$, in units of $[10^{-32}\, \mathrm{cm}^2]$, correspond to 
\begin{align}
-83 \leq \langle{r_{\nu_{\tau}}^{2}}\rangle\,(\mathrm{XnT-ER}) \leq 88 \,,&\quad 
-95 \leq \langle{r_{\nu_{\tau}}^{2}}\rangle\,(\mathrm{XnT-NR}) \leq 35 \,, \\ 
-94 \leq \langle{r_{\nu_{\tau}}^{2}}\rangle\,(\mathrm{P4T-ER}) \leq 99\,, &\quad
-105 \leq \langle{r_{\nu_{\tau}}^{2}}\rangle\,(\mathrm{P4T-NR}) \leq 46 \,,\\ 
-96 \leq \langle{r_{\nu_{\tau}}^{2}}\rangle\,(\mathrm{LZ-ER}) \leq 100 \,.&
\end{align}
These results update those reported in Ref.~\cite{AtzoriCorona:2025xwr} by including the latest ER data from LZ~\cite{LZ:2024zvo} and P4T~\cite{PandaX:2024cic} and by fitting the total number of CE$\nu$NS events from the US2 and paired data from P4T~\cite{PandaX:2024muv}. Instead, considering the spectral analysis for P4T we obtain a weaker constraint of $-120 \leq \langle{r_{\nu_{\tau}}^{2}}\rangle\,(\mathrm{P4T-NR}) \leq 61$ in units of $[10^{-32}\, \mathrm{cm}^2]$. Our results are in good agreement with those reported in Ref.~\cite{Demirci:2025poc}, with our constraint from P4T ER being slightly more stringent than theirs, while the opposite holds for XnT ER data. \\
Although these constraints are still rather broad, the much tighter determinations of the electron and muon flavor charge radii from a global fit of neutrino data ~\cite{AtzoriCorona:2025xwr}, namely
\begin{align}
\langle{r_{\nu_{e}}^{2}}\rangle\,(\text{Global Fit})=&2.2^{+2.4}_{-2.3}\,\times 10^{-32}\, \mathrm{cm}^2\,,\quad \langle{r_{\nu_{\mu}}^{2}}\rangle\,(\text{Global Fit})=-0.19^{+0.55}_{-0.56}\,\times 10^{-32}\, \mathrm{cm}^2\, ,
\label{eq:CR_limit_global fit}
\end{align}
at $1\sigma\, \mathrm{CL}$, can be exploited to improve the precision on the $\tau$ flavor charge radius, when all data are combined together. This combination yields the constraint shown in Fig.~\ref{fig:NCRtau}, corresponding to an updated global fit determination at $1\sigma$ CL of
\begin{equation}
\langle{r_{\nu_{\tau}}^{2}}\rangle\,(\text{Global Fit})=7.5^{+9.5}_{-11.7}\times 10^{-32}\,\mathrm{cm}^2\,.
\end{equation}


\subsection{BSM neutrino electromagnetic properties}
\label{sec:magnetic}
\begin{figure}
    \centering
    \includegraphics[width=0.65\linewidth]{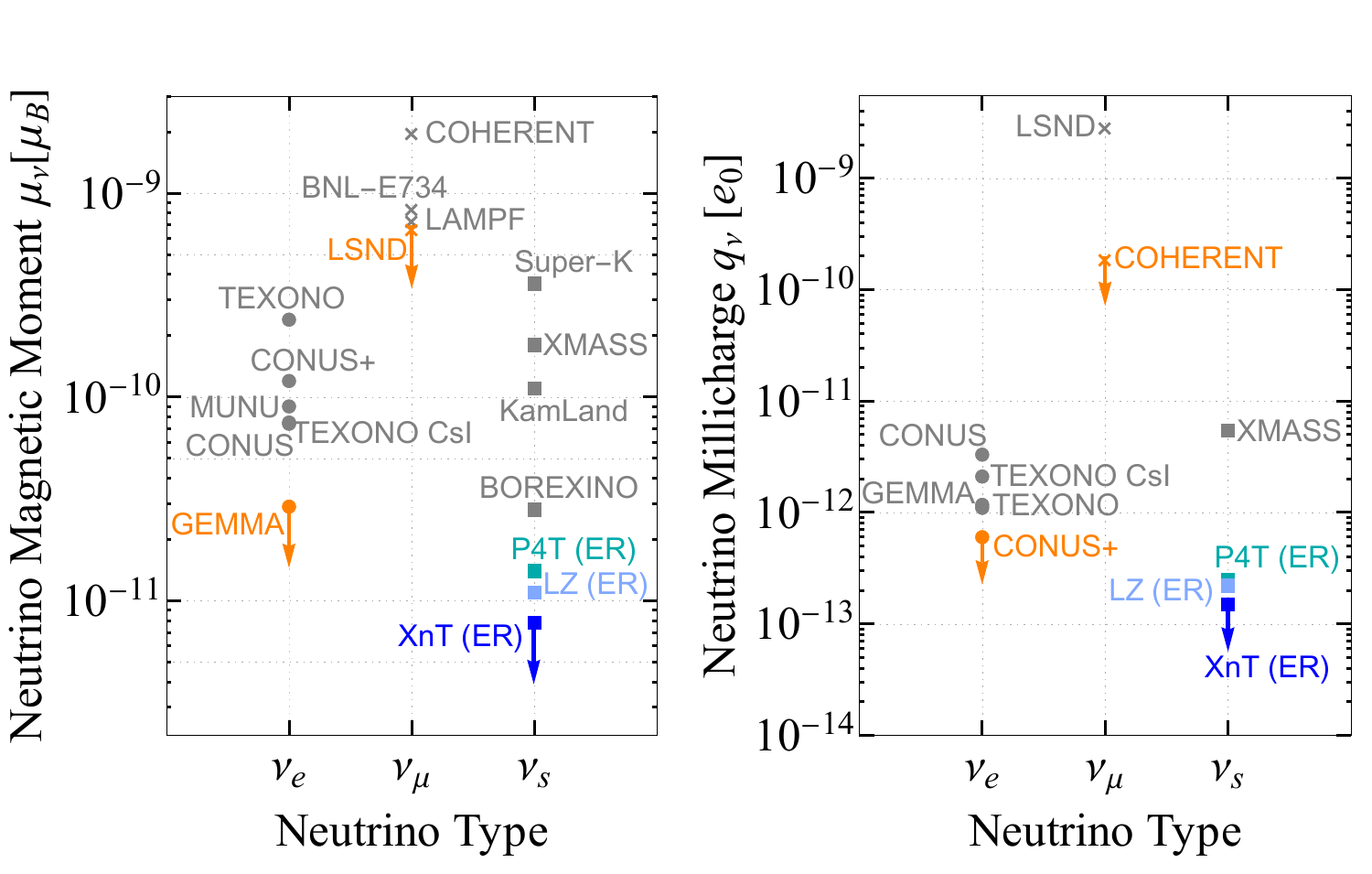}
    \caption{Summary of existing limits at 90\% CL on the neutrino magnetic moment (left) and neutrino millicharge (right) from a variety of experiments~\cite{AtzoriCorona:2022jeb,Beda:2012zz,TEXONO:2006xds,Borexino:2017fbd,Super-Kamiokande:2004wqk,AtzoriCorona:2022qrf,MUNU:2005xnz,Allen:1992qe,Ahrens:1990fp,LSND:2001akn,Giunti:2014ixa,XMASS:2020zke,PhysRevD.110.030001,CONUS:2022qbb, XENON:2022ltv}. The limits are divided into flavor components, and also those on the effective solar parameters are shown. In orange we display the best current limits, which have to be compared to the results obtained from the analysis of ER data from XnT, P4T and LZ carried out in this work.}
    \label{fig:MMandMC}
\end{figure}

When considering beyond the standard model scenarios, additional neutrino electromagnetic properties may arise. In this context, the most investigated one is the neutrino magnetic moment (MM), as its existence may stem from the massive nature of neutrinos~\cite{Giunti:2014ixa,Giunti:2015gga,Giunti:2024gec}.
The MM contribution does not interfere with the SM one, and thus it is accounted for via an additional contribution to the cross section, given by
\begin{equation}
\dfrac{d\sigma_{\nu_{\ell}\text{-}\mathcal{N}}^{\text{MM}}}{d T_\mathrm{nr}}
=
\dfrac{ \pi \alpha^2 }{ m_{e}^2 }
\left( \dfrac{1}{T_\mathrm{nr}} - \dfrac{1}{E_\nu} \right)
Z^2 F_{Z}^2(|\vec{q}|^2)
\left| \dfrac{\mu_{\nu_{\ell}}}{\mu_{\text{B}}} \right|^2 ,
\label{cs-mag}
\end{equation}
for the CE$\nu$NS case, where $\mu_{\nu_{\ell}}$ is the effective MM for an $\ell$ flavor neutrino~\cite{Giunti:2014ixa}
and $\mu_{\text{B}}$ is the Bohr magneton, while for the $\nu$ES process it is
\begin{equation}
\dfrac{d\sigma_{\nu_{\ell}\text{-}\mathcal{A}}^{\text{MM}}}{d T_\mathrm{e}}
=
Z_{\text{eff}}^{\mathcal{A}}(T_{\text{e}}) \dfrac{ \pi \alpha^2 }{ m_{e}^2 }
\left( \dfrac{1}{T_\mathrm{e}} - \dfrac{1}{E} \right)
\left| \dfrac{\mu_{\nu_{\ell}}}{\mu_{\text{B}}} \right|^2.
\label{es-mag}
\end{equation}
The MM contribution depends strongly on the recoil energy, and thus low threshold experiments, such as dark matter detectors, are particularly sensitive to it, especially the $\nu$ES data. Therefore, we report only the results obtained by considering the least-square function in Eq.~(\ref{eq:chi2xe}) for the $\nu$ES datasets\footnote{The constraints from the NR channel have already been determined in Ref.~\cite{DeRomeri:2024hvc}, which result to be approximately 2 order of magnitude weaker than the ER ones.} in terms of the effective solar neutrino magnetic moment $\mu_{\nu_s}$, which at 90\% CL results to be
\begin{align}
\mu_{\nu_s}(\mathrm{XnT-ER})<7.8\times 10^{-12}\mu_B, \\
\mu_{\nu_s}(\mathrm{P4T-ER})<14\times 10^{-12}\mu_B,\\
\mu_{\nu_s}(\mathrm{LZ-ER})<11\times 10^{-12}\mu_B.
\end{align}
These results are also shown in the left panel of Fig.~\ref{fig:MMandMC} together with a compilation of other available measurements~\cite{atzoricorona:2025ygn}.
Our results can be compared with those reported by the experimental collaborations~\cite{XENON:2022ltv,PandaX:2024cic,LZ:2024zvo}; we obtain a slightly stronger constraint in the case of P4T and a slighly weaker one for XnT and LZ. Moreover, in Ref.~\cite{Demirci:2025poc}, the constraint from both P4T and XnT ER data have been derived, showing an excellent agreement with our results.

It is interesting to notice that the constraints set with solar neutrino data are among the strongest available. However, as discussed in Ref.~\cite{Ternes:2025lqh}, the comparison between the effective solar parameter and the different flavor contributions from artificial-source experiments is not trivial, and  therefore it has to be considered as qualitative.

Similarly, the \cenns and the $\nu$ES processes are also sensitive to the possible existence of neutrino electric charges (ECs), also known as millicharges~\cite{Giunti:2024gec}. In contrast to the MM contribution, that of the electric charge adds to SM neutral-current weak interactions, and it can be accounted for via the replacement of the neutrino-proton and the neutrino-electron vector couplings inside the cross sections in Eq.~(\ref{eq:cexsec}) and Eq.~(\ref{eq:ES-cross-section}) with~\cite{Kouzakov:2017hbc,Giunti:2014ixa}
\begin{equation}
g_{V}^{p}\rightarrow 
g_{V}^{p}-\dfrac{ 2 \sqrt{2} \pi \alpha }{ G_{\text{F}} q^2 }
\, q_{\nu_{\ell}}
,\quad 
g_{V}^{\nu_{\ell}}
\to
g_{V}^{\nu_{\ell}} +\dfrac{ 2 \sqrt{2} \pi \alpha }{ G_{\text{F}} q^2 }
\, q_{\nu_{\ell}},
\label{Qech}
\end{equation}
where $q_{\nu_{\ell}}$ is the neutrino EC for a neutrino of flavor $\ell$. Given the extremely low momentum transfer and low-energy thresholds of dark matter experiments, the $q^2$ dependence in the denominator of Eq.~(\ref{Qech}) helps to set stringent constraints, especially considering the case of the $\nu$ES process, because of the much lighter electron mass. Moreover, various studies~\cite{Chen_2017,PhysRevA.25.634,PhysRevA.26.734,Chen:2013lba} have shown that the cross section for a millicharged neutrino at low energies may be further enhanced by more than an order of magnitude due to atomic structure effects not encoded in the effective atomic number $Z_{\rm eff}^{\mathcal{A}}$~\cite{Chen:2014ypv,Hsieh:2019hug}. To improve the treatment, we consider the well established EPA scheme~\cite{Chen:2014ypv,Hsieh:2019hug}. In particular, the EPA cross section for a millicharged ultrarelativistic particle is given by~\cite{Chen:2014ypv,Hsieh:2019hug, AtzoriCorona:2022jeb}
\begin{equation}
    \dfrac{d\sigma_{\nu_\ell}}{d T_\text{e}}\Big\vert_{\rm{EPA}}^{\rm{EC}}=\frac{2\alpha}{\pi}\frac{\sigma_\gamma(T_e)}{T_e}\ln\left[\frac{E_\nu}{m_\nu}\right]q_{\nu_\ell}^2
    \label{eq:EPA},
\end{equation}
where $m_\nu$ represents the neutrino mass, set conservatively to $1\;\rm{eV}$~\cite{PhysRevD.110.030001}, while $\sigma_\gamma(T_e)$ denotes the photoelectric cross section for a real photon, which can be experimentally measured for Xe~\cite{HENKE1993181}. Moreover, in this case we enlarge the systematic contribution to $\sigma_\beta=0.2$ in Eq.~(\ref{eq:chi2xe}) in order to account for additional uncertainties on the EPA approach~\cite{Chen:2014dsa}.
The results at 90\% CL obtained in this analysis for the effective solar neutrino millicharge $q_{\nu_s}$ from ER data are
\begin{align}
-1.4<q_{\nu_s}(\mathrm{XnT-ER})\, [10^{-13}e_0]&<1.7, \\
-2.5<q_{\nu_s}(\mathrm{P4T-ER})\, [10^{-13}e_0]&<2.5,\\
-2.3<q_{\nu_s}(\mathrm{LZ-ER})\,[10^{-13}e_0]&<2.2,
\end{align}
with $e_0$ being the absolute value of the electron charge.
These results are also shown in the right panel of Fig.~\ref{fig:MMandMC} together with a compilation of other available measurements~\cite{AtzoriCorona:2025xgj}. As in the case of neutrino magnetic moments, ER data from dark matter detectors provide the strongest laboratory constraints on the neutrino millicharge. 
Our result from the LZ ER data is slightly weaker than the official one~\cite{LZ:2024zvo}, and in general we find good agreement with the results reported in Ref.~\cite{Demirci:2025poc} for both P4T and XnT, although we find slightly more stringent bounds.\\
It is also important to remark that the constraints on models such as neutrino magnetic moments and neutrino millicharges are strongly dependent on the experimental thresholds and background contributions, along with the corresponding systematics. In our analysis of ER data, as shown in the chi-square function in Eq.~(\ref{eq:chi2xe}), we consider a unique background contribution fluctuating with a single nuisance parameter, which may lead to slightly stronger constraints.


\subsection{Nonstandard interactions and Light Mediators}
\begin{figure}
    \centering
    \subfigure[]{\label{fig:NSI}
    \includegraphics[width=0.49\linewidth]{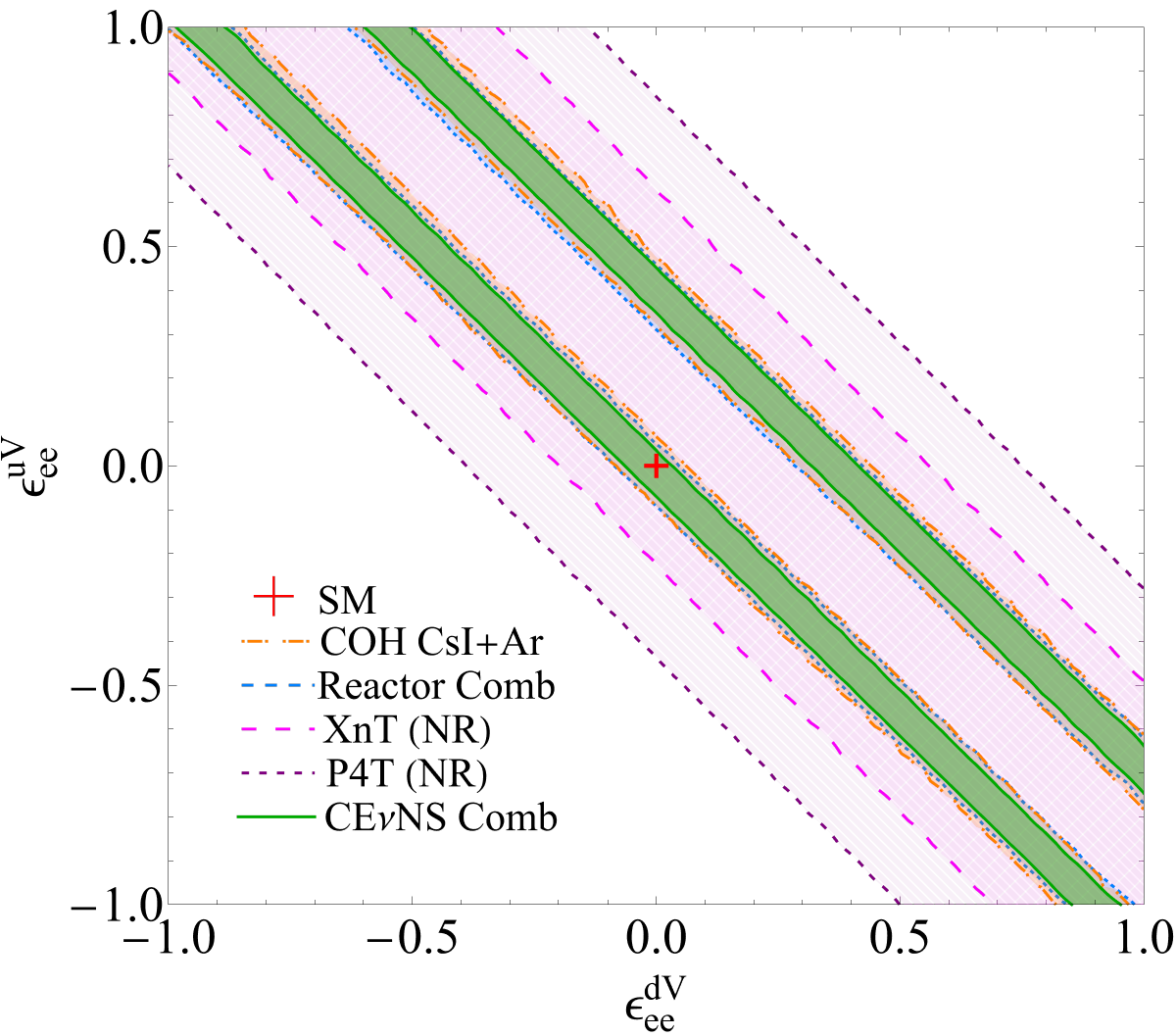}}
    \subfigure[]{\label{fig:LM}
    \includegraphics[width=0.49\linewidth]{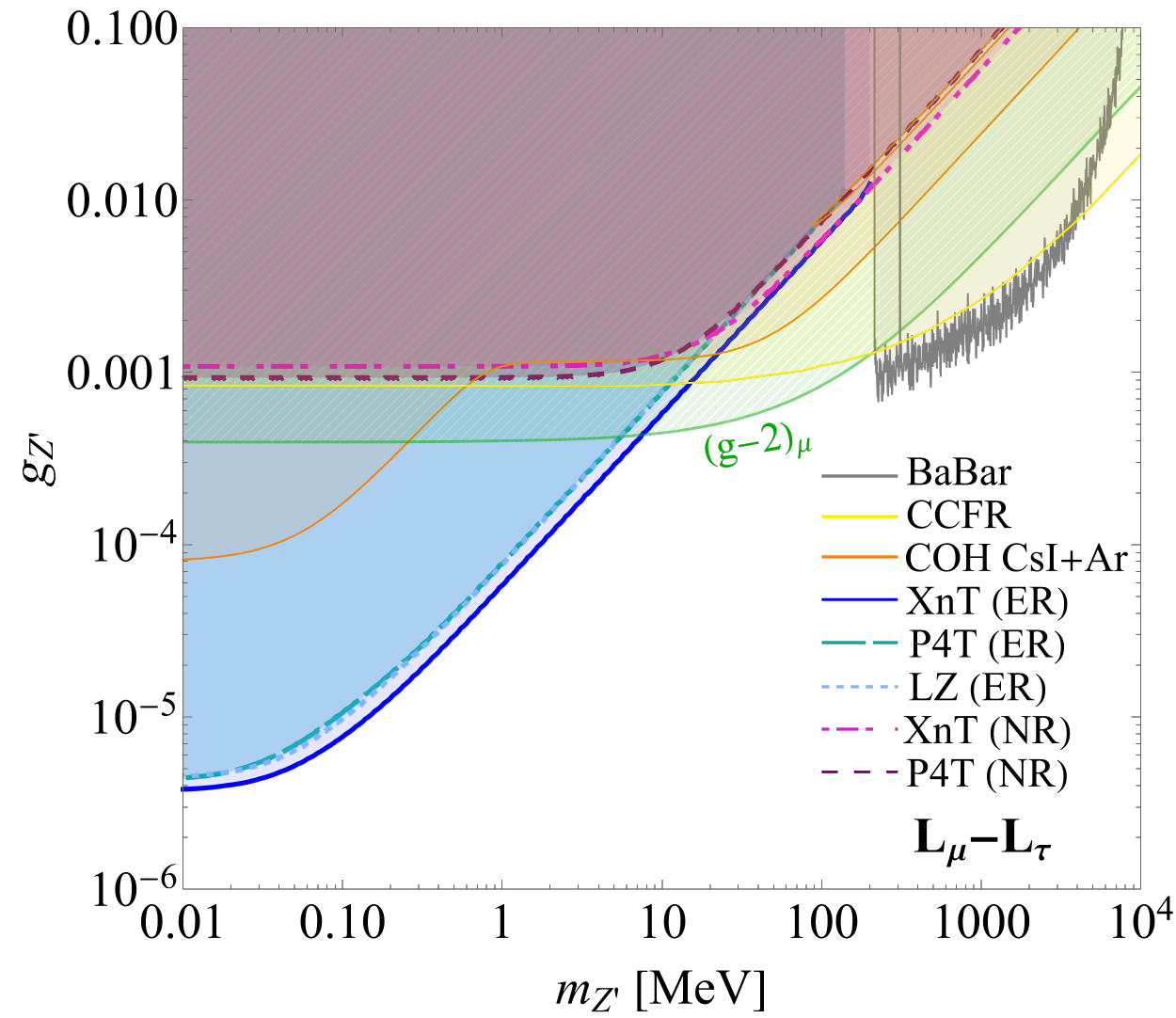}}    
    \caption{Left: Constraints on flavor-preserving NSI at 90\% CL from \cenns solar neutrino data compared to the other \cenns available constraints and the result of a \cenns combined fit (green contour). The red cross indicates the SM solution. Right: Constraints at $2\sigma$ CL on the $L_\mu-L_\tau$ light vector mediator model from ER and NR solar neutrino data compared to other experimental constraints~\cite{BaBar:2016sci,CCFR:1991lpl,Altmannshofer:2014pba,AtzoriCorona:2022moj} and to the limit obtained from the latest $(g-2)_\mu$ experimental result and the lattice-QCD theoretical estimation~\cite{Muong-2:2025xyk,Aliberti:2025beg}.}
\end{figure}

The \cenns and $\nu$ES cross sections may also be modified by the presence of a new massive mediator that couples to SM leptons and quarks, allowing one to test extensions of the SM usually known as light mediators.
A rather general scenario is that of the neutrino non standard interactions, in which a heavy mediator modifies the neutrino-quark coupling by an energy independent contribution. 
Assuming that the neutrino does not change flavor, the full \cenns cross section is obtained by replacing the nuclear weak charge in Eq.~(\ref{eq:weakcharge}) by~\cite{Coloma:2017ncl}
\begin{eqnarray}
Q_{\ell,\mathrm{NSI}}^{V}
=
\left( g_{V}^{p}(\nu_{\ell}) + 2 \varepsilon_{\ell\ell}^{uV} + \varepsilon_{\ell\ell}^{dV} \right)
Z
F_{Z}(|\vet{q}|^2)+\left( g_{V}^{n} + \varepsilon_{\ell\ell}^{uV} + 2 \varepsilon_{\ell\ell}^{dV} \right)
N
F_{N}(|\vet{q}|^2).
\label{Qalpha2}
\end{eqnarray}
In principle one can consider many combinations of NSI parameters, however, in this work we restrict the analysis to a simplified scenario with only two nonzero NSI parameters, namely $\varepsilon_{ee}^{uV}$ and $\varepsilon_{ee}^{dV}$, also known as flavor-preserving scenario, which involves only the electronic neutrino flavor. Given that the NSI depend on the interaction of leptons with quarks, the $\nu$ES cross section is not affected by this type of new physics, and thus we only consider them in the \cenns data analyses. The results obtained using the least-square function in Eq.~(\ref{chi2-NR}) are shown in Fig.~\ref{fig:NSI} at 90\% CL, together with those obtained from other CE$\nu$NS data~\cite{atzoricorona:2025ygn}, and their combination. 
In particular, combining \cenns data on different nuclei, it is possible to improve the constraints profiting from the different neutron and proton number combinations inside the nuclear weak charge. Currently, the contribution from solar neutrino data is only marginal, however, it is expected to become significant in the near future~\cite{Suliga:2020jfa,Beatty:2025bgq} and it will complement the future precision by artificial source experiments~\cite{AtzoriCorona:2025ibl,Chatterjee:2022mmu,COHERENT:2022nrm}.\\
A relevant detail to be addressed in solar neutrino NSI is whether and how they might impact the neutrino oscillation mechanism, the so-called effects in propagation, and not only on the detection. Here, we have neglected such effects, considering canonical neutrino oscillation probabilities~\cite{Borexino:2017fbd,PhysRevD.110.030001}. However, an estimation of such effect has been discussed in Refs.~\cite{Li:2024iij,AristizabalSierra:2024nwf}. In particular, our results are perfectly in agreement with those reported in Ref.~\cite{Li:2024iij} when no effect is considered on the propagation, while they are marginally broader than those where the effect is included.\\
It is also interesting to investigate the scenario in which such novel interactions are generated by a light vector boson, often known as $Z'$, arising from a new gauge symmetry $U(1)'$. In this case, the NSI parameters $\varepsilon$ take the typical form of a propagator~\cite{Cadeddu:2020nbr,AtzoriCorona:2022moj}. Depending on the specific $U(1)'$ model considered, the presence of a new light mediator may arise at tree-level or, if no direct coupling with all of the involved fermions is foreseen, at loop-level through kinetic mixing with the photon, leading in general to a smaller effect with respect to the other popular light mediator models, such as the so-called B-L or universal models~\cite{AtzoriCorona:2022moj,Demirci:2023tui}.

In order to profit from the presence of $\tau$ neutrinos in the solar fluxes, we consider the so-called $L_\mu-L_\tau$ light mediator model (for a more detailed discussion see Refs.~\cite{AtzoriCorona:2022moj,DeRomeri:2024dbv,Demirci:2025qdp,Ghosh:2024cxi,Amaral:2020tga}). This is based on a flavor structure involving only the $\mu$ and $\tau$ flavors. Therefore, no tree-level contribution is present neither for the CE$\nu$NS nor the $\nu$ES scattering processes. The interaction contributes at loop level, through kinetic mixing with the photon, thus effectively intervening as a modification of the neutrino vector coupling with charged particles, i.e. the proton in the CE$\nu$NS case and the electron for the $\nu$ES process. 
In particular, the couplings get modified by
\begin{align}
    g_{V}^{p}\left(\nu_{\ell}\right)\rightarrow g_{V}^{p}\left(\nu_{\ell}\right)+
  \frac{\sqrt{2}\alpha g_{Z^{\prime}}^{2}      \left( \delta_{\ell\mu} \varepsilon_{\tau\mu}(|\vec{q}|) + \delta_{\ell\tau} \varepsilon_{\mu\tau}(|\vec{q}|) \right)}{\pi G_{F} \left(|\vec{q}|^{2}+M_{Z^{\prime}}^2\right)}\,, \quad g_{V}^{\nu_\ell}\rightarrow g_{V}^{\nu_\ell}-
  \frac{\sqrt{2}\alpha g_{Z^{\prime}}^{2}      \left( \delta_{\ell\mu} \varepsilon_{\tau\mu}(|\vec{q}|) + \delta_{\ell\tau} \varepsilon_{\mu\tau}(|\vec{q}|) \right)}{\pi G_{F} \left(|\vec{q}|^{2}+M_{Z^{\prime}}^2\right)},
\end{align}
where $g_{Z'}$ and $M_{Z'}$ are the coupling and mass of the $Z'$ boson, respectively. Here, we have introduced the Kronecker delta $\delta$ to preserve the proper flavor dependences and $\varepsilon_{\tau\mu}(|\vec{q}|)$, which is the one-loop kinetic mixing coupling, given by~\cite{Banerjee:2021laz}
\begin{equation}
    \varepsilon_{\tau\mu}(|\vec{q}|) =
    \int_{0}^{1} x(1-x)
    \ln\left(
    \frac{ m_{\tau}^{2} + x(1-x) |\vec{q}|^{2} }
         {m_{\mu}^{2} + x(1-x) |\vec{q}|^{2} }
    \right)
    d x\, ,
\label{eps}
\end{equation}
where $m_{\tau}$ and $m_{\mu}$ are the tau and muon masses. 
For the energy regime probed by dark matter experiments, $\varepsilon_{\tau\mu}$ is almost constant, because $|\vec{q}| \ll m_\tau$ and $|\vec{q}| < m_\mu$. Therefore the latter can be approximated with $\varepsilon_{\tau\mu}\simeq\ln(m_\tau^2/m_\mu^2)/6$, as done in Refs.~\cite{Altmannshofer:2019zhy,Cadeddu:2020nbr,Bertuzzo:2021opb}.
By considering both ER and NR data and the corresponding least-square functions in Eq.~(\ref{eq:chi2xe}) and Eq.~(\ref{chi2-NR}), we are able to set competitive constraints on the $Z'$ coupling and mass across a large range of the parameter space. Being the momentum transfer much lower for $\nu$ES than CE$\nu$NS, the constraints from ER data are the dominant ones at low masses, while the NR data are more relevant for heavier mediators. 
Similarly to the case of NSI, also for the $L_\mu-L_\tau$ $Z'$ model the oscillation probability could get modified. 
However, given that it contributes at loop level via kinetic mixing, with a coupling proportional to the electric charge, such an effect is expected to be small, given the general neutrality of matter, as also discussed in Ref.~\cite{Amaral:2020tga}.
Our results are shown in Fig.~\ref{fig:LM} compared to the constraints coming from other experimental measurements~\cite{BaBar:2016sci,CCFR:1991lpl,Altmannshofer:2014pba,AtzoriCorona:2022moj}.
They may also be compared to the results reported in Ref.~\cite{Demirci:2025qdp}, where the same model has been studied for the P4T and XnT ER data, finding very similar constraints. In the latter work, other $Z'$ models have also been considered as done in Refs.~\cite{Blanco-Mas:2024ale,DeRomeri:2025nkx,DeRomeri:2024iaw} for the NR data.
In Fig.~\ref{fig:LM}, we compare our results also to the constraints from the recent experimental measurement of the muon anomalous magnetic moment $(g-2)_\mu$~\cite{Muong-2:2025xyk}. For a long time, the latter has shown a significant discrepancy at more than $4\sigma$ CL with respect to the SM prediction~\cite{Aoyama:2020ynm}, motivating the study of such light mediator models as explanations for the experimental result. Indeed, in the presence of a light vector mediator interacting with the muonic flavor, the muon anomalous magnetic moment receives an additional contribution~\cite{Brodsky:1967sr}
\begin{equation}\label{eq:d_amu_th}
\Delta a_\mu^{V} = \frac{g_{Z'}^2}{8\pi^2} \int_{0}^{1} dx \,
\frac{ 2x^2(1-x) }{ x^2 + \left( 1 - x \right) M_{Z'}^2 / m_\mu^2 }.
\end{equation}
However, the latest theoretical calculations, based on the evaluation of the hadronic vacuum polarization from lattice-QCD results~\cite{Aliberti:2025beg}, have now reconciled theory and experiments, and thus, the $(g-2)_\mu$ result leads to a constraint, instead of an allowed strip, in the parameter space, as shown in Fig.~\ref{fig:LM}.

\section{Conclusions}
\label{sec:conclusions}
The recent observation of solar neutrino CE$\nu$NS by the XENONnT and PandaX-4T collaborations marks a significant step forward in the study of neutrino interactions at low energies. Together with the measurements of solar neutrino-electron scattering by the same detectors and the electron recoil result from Lux-Zeplin, these CE$\nu$NS data complement existing CE$\nu$NS and $\nu$ES measurements employing reactor and accelerator neutrinos.
By analysing the available data, we have explored their implications for both SM and BSM physics, with special attention to flavor-dependent scenarios involving the $\tau$ flavor. In particular, we have provided novel determinations of the weak mixing angle, in good agreement with the SM prediction, and improved constraints on neutrino electromagnetic properties such as charge radius, electric millicharge and magnetic moment. In the latter two cases, the electron recoil data, sensitive to the $\nu$ES process, lead to some of the most stringent limits from laboratory experiments.\\ 
Additionally, we have set updated constraints on nonstandard neutrino interactions and on the existence of a hypothetical $L_\mu-L_\tau$ light vector boson mediator. The combination of different target materials with varying proton and neutron numbers, enables a significant improvement over existing bounds.
Moreover, our results show that solar neutrino data from dark matter detectors can contribute substantially to the global analysis of low-energy neutrino data and help in testing the standard model. In particular, solar neutrino data extend and complement the overall picture provided by recent artificial source experiments, allowing searches for possible BSM effects in the $\tau$ sector. Furthermore, solar neutrino CE$\nu$NS data have allowed us to determine the normalization of the $^{8}$B solar neutrino flux, finding good agreement with the GS98 solar model, and to set an upper limit on the $hep$ flux component. Looking ahead, the expected improvement in the precision of these measurements will be crucial for refining these results and probing new physics with even greater sensitivity.

\begin{acknowledgements}
The authors gratefully acknowledge G. Volta for the fruitful discussions about XENONnT data analysis.
\end{acknowledgements}

\bibliography{ref}

\end{document}